\documentstyle{article}
\begin{document}

\title{On the age vs depth and optical clarity \\
of deep ice at South Pole}

\author{The AMANDA collaboration$^{\star}$}

\date{}

\maketitle

\begin{abstract}
The first four strings of phototubes for the AMANDA high-energy neutrino
observatory are now frozen in place at a depth of 800 to 1000 m in ice at the
South Pole.  During the 1995-96 season an additional six strings will be
deployed at greater depths.  Provided absorption, scattering, and refraction of
visible light are sufficiently small, the trajectory of a muon into which a
neutrino converts can be determined by using the array of phototubes to measure
the arrival times of \v{C}erenkov light emitted by the muon.  To help in
deciding on the depth for implantation of the six new strings, we discuss
models of age vs depth for South Pole ice, we estimate mean free paths for
scattering from bubbles and dust as a function of depth, and we assess
distortion of light paths due to refraction at crystal boundaries and
interfaces between air-hydrate inclusions and normal ice. We conclude that the
depth interval 1600 to 1800 m will be suitably transparent for the next six
AMANDA strings and, moreover, that the interval 1600 to 2100 m will be suitably
transparent for a future 1-km$^3$ observatory except possibly in a region a few
tens of meters thick at a depth corresponding to a peak in the dust
concentration at 60 kyr BP.
\end{abstract}

\newpage

\section*{\normalsize \bf INTRODUCTION}

With the creation of AMANDA (the Antarctic Muon and Neutrino Detector Array),
we hope to open a new astronomical window, using neutrinos instead of light
(Lowder and others, 1991; Barwick and others, 1992).  The goal is to image the
trajectories of ultrahigh-energy neutrinos created throughout the universe in
very energetic processes such as take place in quasars and other Active
Galactic Nuclei.  Because the conversion of neutrinos into detectable particles
is extremely small, the observatory must be huge, eventually of order 1 km$^3$
in
volume.  All designs share the same idea:  a three-dimensional array of large
phototubes embedded in a transparent medium will look down at the cone of
\v{C}erenkov light emitted by an upward-moving muon into which a high-energy
neutrino is transformed in passing upward through the Earth.  Measurement of
arrival times of \v{C}erenkov photons at various phototubes enables the
direction of the neutrino to be reconstructed.  In contrast to other
techniques, which propose to build an observatory in a deep lake or the deep
ocean, all of the electronics in the present version of AMANDA sit on the
surface of South Pole ice above cables that simply carry signals up from the
phototubes, and the imaging medium is essentially free of background light such
as that due to radioactivity or bioluminescence.

To minimize cost, the strings of phototubes should be far apart and the spacing
of phototubes on each string should be large.  To reduce background due to
downward- directed muons created in the Earth's atmosphere, the array should be
buried deeply and the phototube spacing should be large in order to give a long
enough lever arm to determine trajectories very accurately.  Our results from
the first four strings, which were deployed during the 1993-94 drilling season,
showed that \v{C}erenkov light from downward muons propagated diffusively due
to scattering from residual air bubbles at the depth interval 800 to 1000 m
where the phototubes were located (Askebjer and others, 1994).  Using a laser
to send pulses of light (500 nm) down optical fibers to various depths, we
measured the distribution of arrival times at phototubes at various distances
from the emitter.  From the excellent fits to a model of random walk with
absorption, we were able to determine separately the absorption length (the
e-folding distance for loss of photons at a particular wavelength) and the
scattering length (the mean free path between scatterers).  The result for the
absorption length at 500 nm, $\lambda_{abs}$ = 59 $\pm$ 3 m, is larger than
that of the
clearest ocean water and far larger than the values estimated for laboratory
ice and for Lake Baikal.  This large value of $\lambda_{abs}$ will permit us
ultimately to
construct a neutrino observatory at lower cost than we originally expected.
The geometrical scattering length on bubbles, $\lambda_{bub}$, was found to be
11 cm at
800 m and to increase to 26 cm at 1000 m.  In order to take maximum advantage
of the information from \v{C}erenkov wavefront arrival times, we plan to
implant
all future strings of AMANDA phototubes at depths sufficiently great that
$\lambda_{bub}$
is greater than about 20 m.

Figure 1 is a sketch showing the location of the first four AMANDA strings and
the proposed location of the remaining six strings.  In this paper we discuss
the factors that must be considered in deciding how deep to implant the
remaining six strings.  Future plans to proceed to construction of a 1
km$^3$-size
neutrino observatory will depend on success in deploying these six strings and
on results of measurements with them.

\section*{\normalsize \bf TIME-SCALE MODEL (AGE VS. DEPTH)}

To estimate the dust concentration as a function of depth, we require a model
of age as a function of depth for South Pole ice.  The most important
ingredient of such a model is the accumulation rate as a function of time and
distance upstream from the site.  Since very little information on accumulation
rate upstream from the South Pole has been published, there is little point in
developing a sophisticated flowline model.  We start with the information in
Table 1 on deep cores in ice and on drilling sites.  The first four rows are
for Greenland sites and the last four are for Antarctic sites.  The last row
includes information about the South Pole site.  No core deeper than 349 m has
been obtained at this site.

Figure 2 displays age vs depth obtained from references at the bottom of Table
1. Wherever possible, we selected data based on absolute determination of age
using stratigraphy (annual layers of high and low $\delta^{18}$O values, of
electrical
conductivity, of dust, or of dissolved impurities due, for example, to
identifiable volcanic eruptions).  In other cases we followed the authors in
associating patterns of $\delta^{18}$O, $^{10}$Be, and dust concentrations from
core to core
with dated events such as the Last Glacial Maximum.  To complete the picture,
we selectively used ages based on ice-flow models reported by the various
authors.  In case of any discrepancies from one author to another, we used the
most recent work or the one least dependent on flow models.  For the Camp
Century core, ages estimated from Fourier spectral analyses of $\delta^{18}$O
profiles
and comparisons with features in the deep sea record differ by a factor $\sim$
2 for
times greater than $\sim$ 10 kyr.  Various reasons have been given for the
irregularities in the Camp Century time-scale near the bottom of the core
(Dansgaard and others, 1982).  We decided to include only the data for depths
more than 280 m above bedrock, for which stratigraphy is said to be accurate to
$\pm$ 3\% (Hammer and others, 1978).

Figure 3 casts the data in Fig. 2 as a universal function in which depth as a
function of age is shown in dimensionless coordinates.  The dimensionless time
is defined as $t^* \equiv t~a(t)/H$, where $t$ is the time in years, $a$ is the
accumulation rate in meters of ice (at density 0.92 g cm$^{-3}$) per yr, and
$H$ is
the ice thickness (m) to bedrock.  The dimensionless depth is defined as $d^*
\equiv d/H$.  Following the practice of most authors, for times $t$ $<$ 15 kyr
BP we
took $a(t)$ to be its average value during the Holocene, and for 15 $\leq$
$t$ $\leq$ 110 kyr BP we took $a(t)$ to be a constant fraction of its Holocene
value.  For
Antarctic sites, the fraction is taken to be 0.75; for Greenland sites the
fraction is taken to be 0.5.  Our adopted values of $a(t)$ are given in columns
7
and 8 of Table 1.  The rationale for assigning a smaller value to $a(t)$ during
the period 15 to 110 kyr BP is that large portions of the time-scales given in
the references are based on ice-flow models for which the authors (with
justification) assumed a smaller value of $a(t)$ during that period.  This
approach should be most successful for sites near ice divides, for which the
horizontal flowrate is small.  Recently more sophisticated models have been
used, in which the accumulation rate has been taken as continuously varying in
a way that can be inferred from the $^{18}$O record or the $^{10}$Be record
less the
spikes.  Since such data are not available for South Pole ice, we have
contented ourselves with the simpler approach using step-function values as
described above.  For the South Pole site, with its large horizontal flowrate,
the accumulation rate far upstream, from which the deep ice originated, is
somewhat smaller than assumed in our simple model (Mosley-Thompson, 1994).  To
convey an idea of the uncertainty, we will present two age vs depth curves for
the South Pole, one of which (McInnes and Radok, 1984) is based on very low
upstream accumulation rates.

Analytical ice flow models of age vs depth have evolved from work by Nye
(1963),
Weertman (1968), and Dansgaard and Johnsen (1969).  On the assumption
of a constant vertical strain-rate with depth, Nye (1963) found the simple
dependence of depth on age:

$$ d = H(1-exp(-at/H)) \eqno{(1)} $$

where $a$ = constant.  In the dimensionless version,

$$ d^* = 1 - exp(-t^*). \eqno{(2)} $$

More refined models take into account the variation of horizontal flowrate,
vertical strain-rate, and temperature with depth; the dependence of ice
plasticity on temperature and crystal texture; the temperature at the
ice-bedrock interface; the flow paths of ice from the surface at various times
in the past to its present depth, which depends on factors such as the distance
of the core from an ice divide; the variation of accumulation rate upstream as
a function of time and of the position from which the ice originated; the slope
of the bedrock; and the slope at the surface.  Little of this information is
available for the South Pole site.

The universal curve in Fig. 3 is for eq. 2.  One sees that for values of
reduced depth $d^*$ greater than about 0.5 the scatter becomes large.  With the
universal curve as input, the solid curve in Fig. 4 shows our prediction of the
age vs depth for the South Pole. The best direct measurements give quite
consistent values of accumulation rate for recent times:  $a$ = 0.074 m/yr from
annual dust layers extending back to 1590 A.D. (Mosley-Thompson, 1994), $a$ =
0.072 m/yr based on visible stratigraphy in an inclined snow mine for the
period 1950 to 1770 A.D. (Giovinetto, 1964), and 0.073 m/yr based on a stake
network and beta-activity measurements over much of Antarctica (Young and
others, 1982).  The solid curve in Fig. 4 is for $a(t)$ = 0.073 m/yr back to 15
kyr BP and 0.055 m/yr before 15 kyr BP.

One attempt to predict age vs depth for ice at the South Pole has been
published (McInnes and Radok, 1984).  The dashed curve in Fig. 4 shows their
model, which agrees with ours at depths down to $\sim$ 1300 m, corresponding to
an
age of $\sim$ 33 kyr BP.  For greater depths they inferred a much more rapid
increase
in age with depth than we do, because they assumed an abrupt decrease in
accumulation rate from $\sim$ 0.088 m/yr at the Pole to a constant value of
$\sim$ 0.02
m/yr for ice more than $\sim$ 100 km upstream, which reaches the Pole at times
greater than $\sim$ 30 kyr BP.  Mosley-Thompson's recent measurements of
accumulation rate at Plateau Remote, near the Pole of Relative Inaccessibility,
gave values about twice as large as assumed by McInnes and Radok.  The two
curves in Fig. 4 give, with high probability, upper and lower limits on the
true values.

The irregular curve along the abscissa gives the dust flux at South Pole in
arbitrary units as a function of time, inferred from measurements made on an
ice core at Vostok (See later section for discussion).

\section*{\normalsize \bf DEPTH DEPENDENCE OF SCATTERING FROM BUBBLES}

A quantity of great relevance to the AMANDA project is the depth-dependence of
$\lambda_{bub}$, the mean free path for scattering of the visible component of
\v{C}erenkov radiation off of air bubbles.  For smooth spherical bubbles of rms
radius r and number concentration n, $\lambda_{bub} = (n{\pi}r^{2})^{-1}$.  For
purposes of
discussion, the polar ice can be divided into three depth regimes:  At depths
shallower than the closeoff depth (typically $\sim$ 60 m), the firn is porous
to air.
At intermediate depths, bubbles exist at an approximately constant number
concentration and with mean bubble volume that decreases as hydrostatic
pressure increases.  At great depths, bubbles transform into a more stable
phase consisting of air hydrate crystals.

Bubbles are formed at the "closeoff" depth at which the firn changes into ice.
The volume fraction of air trapped in the ice at pore closeoff, $V_c$, is given
experimentally (Raynaud and Lebel, 1979) by

$$ V_c {\rm (cm^3/g~of~ice)} = (2 \times 10^{-4} - 0.015/T(K)) P{\rm (mbar)}
\eqno{(3)} $$

$V_c$ depends on atmospheric pressure and thus on altitude at the time of
closeoff, and depends weakly on temperature.  For Vostok, $V_c$ = 0.084
cm$^3$/g,
whereas for Byrd, at an altitude nearly 2000 m lower than Vostok, $V_c$ = 0.114
cm$^3$/g.   The South Pole, with altitude intermediate between Vostok and Byrd,
should have an intermediate value of $V_c$.

At z$_c$, the ice-equivalent depth for pore closeoff, $\rho V_{c}/n = 4{\pi}
r^{3}/3$ if all the
gas is in bubbles (where $\rho$ = mass density).  The number concentration at
pore
closeoff depends on temperature, dust concentration, and other climatic
factors, but is roughly similar for Vostok and Byrd, the two cores in which
detailed studies have been made of bubbles. Since the ice density is roughly
constant below the firn layer and the ice temperature does not change much in
the first 1000 m, the perfect gas law leads to expressions for the depth
dependence of the rms bubble radius

$$ r = ({ {3\rho V_c z_c} \over {4 \pi n z} })^{1/3} \eqno{(4)} $$

and of 1/$\lambda_{bub}$:

$$ 1/\lambda_{bub} = ({ {3\rho V_c z_c n^{1/2} \pi^{1/2}} \over {4z} })^{2/3}
\eqno{(5)} $$

The latter is shown by the solid curve in Fig. 5.  The figure also shows data
for 1/$\lambda_{bub}$ at Byrd, Vostok, and the South Pole, derived from
measurements by
Gow and Williamson (1975) on a Byrd core, by Barkov and Lipenkov (1984) on a
Vostok core, and on laser calibration data in situ by the AMANDA collaboration
(Askebjer and others, 1994).  The vertical dashed lines for the in situ
measurements are put in to remind the reader that the scattering length
inferred from the data depends on the angular distribution of scattered light:
the lower triangles correspond to the assumption of isotropic scattering from
bubbles with rough surfaces, which might come about, for example, if air
hydrate crystals were nucleating at various points on the surfaces) and the
upper triangles correspond to the assumption of forward-peaked scattering from
smooth, spherical bubbles.  (The true values should lie somewhere in between.)
The curve in Fig. 5 is normalized to the data at depths less than 700 m by
choosing z$_c$ = 15 m.

At depths from 100 to nearly 700 m the data for Byrd and Vostok follow the
hydrostatic pressure curve rather well.  At greater depths the inverse
scattering length decreases rapidly with depth, due to a decrease in n brought
about by a phase transition.  A quantitative diffusion-growth model for the
conversion of bubbles into air hydrate crystals (Price, 1994) both fits the
data for 1/$\lambda_{bub}$ as a function of depth at Vostok and Byrd and
predicts the
effective disappearance of bubbles at a depth of $\sim$ 1450 m in South Pole
ice. The
dotted curves in Fig. 5 show the results of that calculation for Vostok and
Byrd and the dashed curve shows the prediction for the South Pole.  In the next
section we  discuss further the phase transition from the two-phase system (ice
+ bubbles) into the two-phase system (ice + air hydrate).

\section*{\normalsize \bf CONVERSION OF BUBBLES TO AIR HYDRATE CRYSTALS}

Miller (1969) first showed that, at sufficiently high pressure, the system (air
bubbles + ice) become unstable against the transformation into a two-phase
system consisting of normal ice + crystals of the cubic clathrate ice structure
in which O$_2$ and N$_2$ molecules occupy a fraction of the clathrate cages
(experimentally, about 80\%).  In Fig. 6 the dashed curves (Miller, 1969) give
the dissociation pressures as a function of ice temperature for nitrogen
hydrate and "air hydrate".  Within the hatched region between the two curves
both bubbles and hydrate crystals should co-exist.  On the same graph we show
the temperature as a function of depth for some cores for which either
measurements or calculations exist.  The profile at South Pole is based on a
model with T fixed at -50$^{\circ}$ C at the surface and at the
pressure-melting
temperature at bedrock.  Seven months after completion of hot-water drilling,
temperatures measured at 800 to 1000 m with thermistors imbedded during the
AMANDA operation had dropped to within $\sim$ 0.3$^{\circ}$ C of the model
temperatures.

With polarized light, air hydrate crystals have been observed in a number of
core sections.  Shoji and Langway (1982; 1987) studied air hydrate crystals in
cores from Dye 3, Camp Century, and Byrd Station. Uchida and others (1994)
recently made quantitative measurements of size, shape, and concentration of
such crystals as a function of depth in Vostok cores.  In Fig. 6 we have
indicated on the depth vs temperature curves the depths at which air hydrate
crystals first appear, as well as the depths at which air bubbles are no longer
present.  Shoji and Langway (1987) pointed out that the depth at which air
hydrate crystals first appear agrees reasonably well with the depths
corresponding to the dissociation pressures calculated for air hydrate by
Miller (1969).  To explain why air hydrate crystals are found at depths about
100 m above the depth corresponding to the equilibrium phase boundary at Byrd
(see our Fig. 6), Craig and others (1993) pointed out that the hydrate crystals
had probably formed at a greater depth just upstream of Byrd and had remained
metastable following vertical advection of ice through the local phase-
boundary depth on the way to their present location at Byrd.  The advection is
attributed to large-scale irregularities in the topography of the bedrock.
Since the bedrock topography upstream of South Pole would not give rise to such
upwelling, we predict that air hydrates will appear in increasing concentration
at depths exceeding $\sim$ 500 m at the South Pole.

Price (1994) showed that the broad transition zone in depth for coexistence of
both bubbles and air hydrate crystals in ice cores was due not to a nucleation
barrier but rather to the slowness of diffusion of water molecules through a
spherical shell of air hydrate crystal forming on bubble walls.  His model
exploited the fact that, whereas the activation energy for diffusion through
normal ice is only 0.57 eV, the activation energy for diffusion through air
hydrate is much higher, 0.9 eV (Uchida and others, 1992), which is able to
account for the very slow growth rate of air hydrate crystals.

In Table 1, column 9 gives the depths at which air bubbles in various cores are
no longer seen.  The same data are displayed in Fig. 6.  For the South Pole
site, we adopt the diffusion model that predicts essentially complete
conversion of bubbles into air hydrate crystals at depths greater than 1450 m.

\section*{\normalsize \bf DUST VS DEPTH}

In the absence of bubbles, insoluble dust particles with diameters $\sim$ 0.1
to $\sim$ 10
$\mu$m make the most important contribution to scattering of visible light in
the ice.  The concentration of dust and soluble impurities in the ice has been
shown to be directly related to the concentration of atmospheric aerosols.  The
size distribution is log normal, with typical modal diameter ranging from 0.5
$\mu$m (in Dome C: Royer and others, 1983) to 0.6 $\mu$m (in Vostok: De Angelis
and others, 1984).  The dust concentration in ice cores shows both an annual
variation and a long-term anticorrelation with the temperature inferred from
$\delta^{18}$O measurements.  For example, as seen in Table 2, all deep cores
show an
increase in dust concentration at depths corresponding to the Last Glacial
Maximum -- by one order of magnitude for Antarctica and by two orders of
magnitude for Greenland.  The vertical dust flux increases as global
temperature decreases, both because of increased continental aridity and
because of higher wind velocity.  (The flux is roughly proportional to the
desert area of nearby continents and to the cube of wind velocity.)  For a
given annual dust flux the concentration deposited in ice is inversely
proportional to snow accumulation rate and is roughly constant for the entire
Antarctic continent (Petit and others, 1990).

In Fig. 4, the irregular curve shows the dust flux as a function of time at
Vostok inferred from measurements on an ice core (Petit and others, 1990).  By
comparing this curve with the two models of age vs depth on the same graph, we
obtain estimates of the depths at the South Pole at which maxima and minima in
the dust concentration occur.  For example, both of the two models predict that
the largest peak, corresponding to the Last Glacial Maximum, will be found at a
depth of $\sim$ 1000 m.  The next maximum, at $\sim$ 60 kyr BP, is predicted to
occur at
$\sim$ 1950 m with the AMANDA model shown by the solid curve and at $\sim$ 1500
m with the
McInnes-Radok model (dashed curve).

Taking together the requirements that all bubbles must have converted into air
hydrate crystals (which dictates that the detectors be deeper than 1450 m),
that depths with high dust concentration be avoided, and that depths of large
horizontal shear ($>$ 2100 m) be avoided, the two models lead to the following
constraints:

\begin{list}{}{}
\item[$\bullet$]With the "AMANDA" model given by the solid curve, the
longest scattering lengths are to be found at 1450 m to 1850 m.
\item[$\bullet$]With the McInnes-Radok model given by the dashed curve, the
longest scattering lengths are to be found at 1600 to 2100 m.
\end{list}

Thus, with either model a $\sim$ 250-m-long vertical string of detectors would
encounter a low dust concentration at 1600 to 1850 m.  In planning for an
eventual 1-km$^3$ array it would be useful to use the pulsed laser technique to
measure lscatt as a function of depth throughout the region of possible
interest from 1450 to 2100 m.  As the AMANDA observatory expands beyond the set
of ten strings envisaged in Fig. 1, additional strings can be implanted so as
to avoid the dust peak at 60 kyr BP.

We next estimate the mean free path for light scattering from dust at the South
Pole for the maximum and minimum values of the dust concentration.  We use data
collected in Table 2, which gives our estimates of the dust concentrations in
various cores at depths corresponding to three times:  the Holocene (from the
present back to $\sim$ 13 kyr BP), for which the dust concentration has been
roughly
constant and quite low; the Last Glacial Maximum, at $\sim$ 18 kyr BP; and a
time $\sim$ 40
kyr BP, typical of the region of minimal dust concentration.  Data at the South
Pole exist only for a depth of 100 to 349 m (Mosley- Thompson, 1994).  To
predict the concentration at 40 kyr BP at South Pole, we note that the ratio of
the concentration at 40 kyr BP and during the Holocene is $\sim$ 1.5 for Byrd,
2.6
for Dome C, and 2.9 for Vostok.  We assume that the average of these three
values, a factor 2.3, holds at South Pole.  Noting that the ratio of
concentrations at LGM and during the Holocene is 10, 17, and 12 for Byrd, Dome
C, and Vostok respectively, we adopt a factor $\sim$ 13 for the South Pole.
Our
predictions appear in row 6 of Table 2.

We estimate the mean free path for light scattering from dust in two ways.  The
more direct approach uses the data in Table 3 on the diameter distribution of
dust measured in an ice core in the depth interval 100 to 349 m at South Pole.
(No deeper core has yet been obtained.)  Column 2 gives the numbers of
particles per ml and column 3 gives the total cross-sectional area per ml
assuming $\pi r^2$ per particle.  Summing the entries gives $7 \times 10^4$
particles
with diameter $>$ 0.1 $\mu$m and total cross-sectional area per ml of
$7 \times 10^{-5}$ cm$^{-1}$.  (For a log-normal size distribution
peaked at 0.5 $\mu$m, the area/ml
below 0.1 $\mu$m (Gayley and Ram (1985) is negligible and the area/ml above 2
$\mu$m (Mosley-Thompson, 1994) is as given in Table 3.)  The reciprocal of the
latter quantity is the mean free path, $\lambda_{scat}$, which is
seen to be $\sim$ 140 m.
Assuming that the relative concentrations in the different diameter intervals
do not change with depth, we can scale the mean free path to different times,
leading to the prediction that $\lambda_{scat} \sim$ 11 m at the Last Glacial
Maximum and
$\lambda_{scat} \sim$ 60 m at a depth corresponding to 40 kyr BP.

The second approach is based on the study of light scattering and absorption by
dust suspended in a Dome C core.  Royer et al. (1983) used a gonionephelometer
to measure the intensity of light at 546 nm scattered at angles from
15$^{\circ}$ to
150$^{\circ}$ in freshly melted ice.  Using Mie theory to fit the angular
distribution
of scattered light, they derived average values of the scattering coefficient
$b$
= 3.46 $\pm$ 0.2 $\times$ $10^{-2}$ m$^{-1}$ for Holocene samples, $b$ =
10 $\pm$ 2.5 $\times$ 10${-2}$ m$^{-1}$
for LGM samples, and $\sim$ 5 $\times$ 10$^{-2}$ m$^{-1}$ for samples near the
bottom of the
core (corresponding to $\sim$ 30 kyr BP).  From these results we infer a mean
free
path $\lambda_{scat}$ = 20 m for dust near the bottom of the core at Dome C.
Scaling with
accumulation rate, we infer $\lambda_{scat}$ = 40 m for dust at 1500 m at South
Pole.
Comparison of this result with that obtained by the first approach provides an
indication of the uncertainty in the value of $\lambda_{scat}$.

\section*{\normalsize \bf REFRACTION BY AIR HYDRATE CRYSTALS}

In the last section we concluded that in the depth interval 1600 to 1850 m at
the South Pole all bubbles will have transformed into air hydrate crystals and
$\lambda_{scat}$ will be of order 40 to 60 m, independent of the age vs depth
model.
Because the air hydrate crystals have dimensions very large relative to the
wavelength of light, \v{C}erenkov light will undergo refraction rather than
scattering at interfaces between air hydrate and hexagonal ice crystals.  The
quantitative effect of the hydrate crystals on light transmission will depend
on their size, shape, concentration, and refractive index relative to that of
the normal ice in which they are imbedded.  Using X-ray diffraction, Hondoh and
others (1990) verified the cubic clathrate nature of natural air hydrate
crystals in a Dye 3 deep core. From the Raman intensities of the stretching
modes of N$_2$ and O$_2$, Nakahara and others (1988) inferred a composition
ratio of
N$_2$ to O$_2$ to be 1.6 - 1.9 in the clathrate structure of these same hydrate
crystals.  Shoji and Langway (1982; 1987) made optical studies and did
laboratory experiments on air hydrate crystals in ice cores from Dye 3, Camp
Century, and Byrd. Uchida (1994) determined that the ratio of the refractive
index of natural air hydrate in ice core samples relative to the refractive
index of hexagonal ice is 1.004.  Uchida and others (1994) recently made an
extensive study of the depth profiles of the shape, volume concentration,
number concentration, and mean volume per crystal of air hydrate crystals in
Vostok ice-core samples.  They found a typical concentration of air hydrate
crystals of $\sim$ 500 cm$^{-3}$, a typical diameter of $\sim$ 100 $\mu$m, and
a predominance of
spherical shapes, from which we infer a typical mean path of $\sim$ 25 cm for
encounter of a light ray with such crystals.  To reach a phototube at a
distance of, say, 25 m requires $\sim$ 10$^2$ traversals of such crystals.  For
a
typical angle of incidence of 45$^{\circ}$ and 10$^2$ encounters, and taking
into account
refraction at both entrance and exit with n = 1.004, the net lateral deflection
due to a random walk about the initial direction would be only $\sim$ 5 cm over
the
25 m pathlength. Thus, because of the nearly perfect match of refractive index
to that of hexagonal ice, air hydrate crystals have virtually no effect on the
imaging of muon trajectories.

\section*{\normalsize \bf REFRACTION AT ICE CRYSTAL BOUNDARIES}

With its hexagonal crystal structure, ice has a refractive index along the
c-axis that is larger than that in the basal plane by a factor of 1.001.
Assuming a mean crystal size of 4 mm, the same as measured in the Vostok core
at a depth of 1500 m (Lipenkov and others, 1989), a light ray in South Pole ice
would refract $\sim$ 6000 times in traversing a distance of 25 m.  To make the
most
conservative estimate, we assume a random distribution of c-axes with a typical
angle of 45$^{\circ}$ between c-axes in adjacent crystals.  Then the net
deflection due
to a random walk of 6000 steps would be only $\sim$ 0.2 cm over the 25 m
pathlength.
Again, we are helped by the very small ratio of refractive indices across a
crystal boundary.

\section*{\normalsize \bf CONCLUSIONS}

\newcounter{pointnum}

\begin{list}{\arabic{pointnum}.}{\usecounter{pointnum}}
\item Almost certainly the true age vs depth relationship lies between the two
curves in Fig. 4.  The AMANDA model (solid curve) probably underestimates ages
at great depths because the choice of $a(t)$ = 0.055 m/yr during the interval
15
to 110 kyr BP is higher than recent values estimated by Mosley-Thompson (1994).
 The flowline model of McInnes- Radok (dashed curve) overestimates the ages at
depths greater than $\sim$ 1300 m because of the very small assumed value
$a(t)$ = 0.02
m/yr at distances more than $\sim$ 180 km upstream.
\item A diffusion-growth model that fits Vostok and Byrd core data predicts
that
the transformation of bubbles into air hydrate crystals at the South Pole will
be complete at depths below $\sim$ 1450 m.
\item Using data on the size distribution of dust in a shallow South Pole core
(100 to 349 m) together with systematics of depth profiles of dust
concentrations at other sites, we estimate a mean free path of $\sim$ 40 to
$\sim$ 60 m for
scattering of light from dust in the depth interval 1600 to 1850 m at South
Pole.  In this depth interval both age models give the same result.  At the
depth corresponding to the peak in the dust concentration at $\sim$ 60 kyr BP,
the
scattering mean free path is estimated to drop to $\sim$ 13 to 20 m.  Such
short mean
free paths are confined to a vertical layer a few tens of meters thick located
somewhere between 1500 and 2000 m depending on the age vs depth model.
\item Taking into account the age vs depth relationship, the depth profile of
the dust concentration for the deep Vostok core, the expected depth for bubble
disappearance (1450 m), and the depth ($\sim$ 2500 m) at which the horizontal
ice flow rate is thought to change rapidly with depth (Koci, 1994), we conclude
that
\begin{list}{}{}
\item[$\bullet$]the next six AMANDA strings should be implanted at a depth of
1600 to 1850 m;
\item[$\bullet$]the future 1-km$^3$ observatory could extend vertically from
1600 to 2100 m, with a horizontal area of $\sim$ 2 km$^2$.  Once the layer of
high dust concentration has been located with the pulsed laser technique, one
could simply exclude data recorded in phototubes in this layer a few tens of
meters thick.
\end{list}
\item Due to the small difference in refractive index between air hydrate and
normal hexagonal ice crystals, refraction at the interfaces between the two
types of crystals will have a negligible effect on the trajectories of
\v{C}erenkov light rays emitted by muons in the ice.
\item Similar reasoning leads to the conclusion that, due to the small
difference
in refractive index parallel to and perpendicular to the c-axis in normal
hexagonal ice, refraction at the interfaces between randomly distributed ice
crystals will have a negligible effect on the trajectories of \v{C}erenkov
light rays emitted by muons.
\end{list}

\section*{\normalsize \bf ACKNOWLEDGMENTS}

We are indebted to Bruce Koci and the entire PICO organization for their
excellent support with the ice drilling.  This work was supported in part by
the National Science Foundation, the K. A. Wallenberg Foundation, the Swedish
Natural Science Research Council, the G. Gustafsson Foundation, Swedish Polar
Research, and the Graduate School of the University of Wisconsin, Madison.  We
thank R. B. Alley, S. P. Davis, P. Duval, J. Fitzpatrick, A. J. Gow, T. Hondoh,
B. Koci, V. Ya. Lipenkov, S. L. Miller, E. Mosley-Thompson, R. G. Pain, J. M.
Palais, J. R. Petit, M. Ram, C. F. Raymond, and T. Uchida for helpful
discussions.

\newpage

$\star$ Members of the AMANDA Collaboration are P. Askebjer$^{\star}$, S.
Barwick$^{\dag}$, L. Bergstr\"{o}m, A. Bouchta$^{\star}$, S. Carius$^{\ddag}$,
A.
Coulthard$^{\S}$, K. Engel$^{\S}$, B. Erlandsson$^{\star}$, A.
Goobar$^{\star}$, L.
Gray$^{\S}$, A. Hallgren$^{\ddag}$, F. Halzen$^{\S}$, P. O. Hulth$^{\star}$, J.
Jacobsen$^{\S}$, S. Johansson$^{\star\P}$, V. Kandhadai$^{\S}$, I.
Liubarsky$^{\S}$, D. Lowder$^{\&}$, T. Miller$^{\&\star\star}$, P.
Mock$^{\dag}$, R.
Morse$^{\S}$, R. Porrata$^{\dag}$, P. B. Price$^{\&}$, A. Richards$^{\&}$, H.
Rubinstein$^{\ddag}$, J. C. Spang$^{\S}$, Q. Sun$^{\star}$, S. Tilav$^{\S}$, C.
Walck$^{\star}$, and G. Yodh$^{\dag}$.  For communication regarding this
manuscript, correspond with P. B. Price (e-mail address price@lbl.gov).

\bigskip

\begin{list}{}{}
\item $^{\star}$Stockholm University, Sweden
\item $^{\dag}$University of California, Irvine, CA
\item $^{\ddag}$Uppsala University, Sweden
\item $^{\S}$University of Wisconsin, Madison, Wisconsin
\item $^{\&}$University of California, Berkeley, CA
\item $^{\P}$Currently at J\"{o}nk\"{o}ping University, Sweden
\item $^{\star\star}$Currently at Bartol Research Institute, Delaware
\end{list}

\newpage

\section*{\normalsize \bf REFERENCES}

\begin{list}{}{}
\item{Askebjer, P., and others.  1994.  {\sl Science}, in press.}
\item{Barkov, N. I., and V. Ya. Lipenkov.  1984.  {\sl Mat. Glyatsiol. Issled.}
{\bf 51}, 178-186. }
\item{Barwick, S., F. Halzen, D. Lowder, T. Miller, R. Morse, P. B. Price, and
A.
Westphal. 1992.  {\sl J. Phys. G: Nucl. Part. Phys.} {\bf 18}, 225-247. }
\item{Craig, H., H. Shoji, and C. C. Langway, Jr.  1993.  {\sl Proc. Natl.
Acad.
Sci.} {\bf 90}, 11416. }
\item{Dahl-Jensen, D.  1989.  {\sl Ann. Glaciology} {\bf 12}, 31-36. }
\item{Dahl-Jensen, D., and S. J. Johnsen.  1986.
{\sl Nature} {\bf 320}, 250-252. }
\item{Dansgaard, W., and S. J. Johnsen.  1969.  {\sl Journal of
Glaciology} {\bf 8} (53), 215-223. }
\item{Dansgaard, W., H. B. Clausen, N. Gundestrup, C. U. Hammer, S. F. Johnsen,
P. M.
Kristinsdottir, and N. Reeh.  1982.  {\sl Science} {\bf 218}, 1273-1277. }
\item{Dansgaard, W., S. J. Johnsen, H. B. Clausen, D. Dahl-Jensen, N.
Gundestrup, C.
U. Hammer, C. S. Hvidberg, J. P. Steffensen, A. E. Sveinbj\"{o}rnsdottir, J.
Jouzel, and G. Bond.  1993.  {\sl Nature} {\bf 364}, 218-220. }
\item{De Angelis, M., M. Legrand, J. R. Petit, N. I. Barkov, Ye. S.
Korotkevich, and
V. M. Kotlyakov.  1984.  {\sl Journal of Atmospheric Chemistry } {\bf 1},
 215-239. }
\item{Gayley, R. I., and M. Ram. 1985.  {\sl Journal of Geophysical
Research } {\bf 90}, 12921-12925. }
\item{Giovinetto, M. B.  1964.  {\sl Antarctic Research Series
(American Geophysical Union)}, {\bf 2}, 127-155. }
\item{Gow, A. J., and T. Williamson.  1975.  {\sl Journal of
Geophysical Research} {\bf 80}, 5101-5108. }
\item{Hammer, C. U., H. B. Clausen, W. Dansgaard, N. Gundestrup,
S. J. Johnsen, and N. Reeh.  1978.  {\sl Journal of Glaciology} {\bf  20},
3-26. }
\item{Hondoh, T., H. Anzai,  A. Goto, S. Mae, A. Higashi, and C. C. Langway,
Jr.
1990. {\sl Journal of Inclusion Phenomena and Molecular Recognition
in Chemistry} {\bf 8}, 17-24. }
\item{Ikeda, T., T. Uchida, and S. Mae.  1993.  {\sl Proc.
NIPR Symp. Polar Meteorol. Glaciol.} {\bf 7}, 14-23. }
\item{J. Jouzel et al.  1993.  {\sl Nature} {\bf 364}, 407-412. }
\item{Kittel, C. and H. Kroemer.  1980.  \underline{Thermal Physics},
second edition (W. H. Freeman, San Francisco). }
\item{Koci, B.  1994.  Private communication. }
\item{Lipenkov, V. Ya., N. I. Barkov,  P. Duval, and P.
Pimienta.  1989. {\sl Journal of Glaciology} {\bf 35}, 392-398. }
\item{Lorius, C., L. Merlivat, J. Jouzel, and M. Pourchet.  1979.
{\sl Nature} {\bf 280}, 644-648. }
\item{Lorius, C., D. Raymond, J. R. Petit, J. Jouzel, and
L. Merlivat.  1984.  {\sl Ann. Glaciology} {\bf 5}, 88-94. }
\item{Lorius, C., J. Jouzel, C. Ritz, L. Merlivat, N. I. Barkov,
Y. S. Korotkevich, and V. M. Kotlyakov.  1985.  {\sl Nature} {\bf 316},
591-596. }
\item{Lowder, D., T. Miller, P. B. Price, A. Westphal, S. W. Barwick, F.
Halzen, and
R. Morse.  1991.  {\sl Nature} {\bf 353}, 331-333. }
\item{McInnes, B. and U. Radok.  1984.  {\sl Antarctic Journal
of the U. S.} {\bf 19} (1), 10-12.}
\item{Miller, S. L. 1969.  {\sl Science} {\bf 165}, 489-490. }
\item{Mosley-Thompson, E.  1994.  Private communication. }
\item{Mosley-Thompson, E., and L. G. Thompson.  1982.
{\sl Quaternary Research} {\bf 17}, 1-13. }
\item{Nye, J. F.  1963.  {\sl Journal of Glaciology} {\bf 4} (36),
785-788. }
\item{Nakahara, J., Y. Shigesato, A. Higashi, T. Hondoh, and C. C. Langway, Jr.
1988. {\sl Philosophical Magazine} {\bf 57}, 421-430. }
\item{Palais, J. M., M. S. Germani, and G. A. Zielinski.  1992.
{\sl Geophysical Research Letters} {\bf 19} (8), 801-804. }
\item{Petit, J. R.,  M. Briat, and A. Royer.  1981.
{\sl Nature} {\bf 293}, 391-394. }
\item{Petit, J. R., L. Mounier, J. Jouzel, Y. S. Korotkevich,
V. I. Kotlyakov, and C. Lorius. 1990.  {\sl Nature} {\bf 343}, 56-58. }
\item{Price, P.B. 1994.  {\sl Science}, in press.}
\item{Raynaud, D., and B. Lebel.  1979.  {\sl Nature} {\bf 281}
 (5729), 289-291. }
\item{Ritz, C.  1989.  {\sl Ann. Glaciology} {\bf 12}, 138-144. }
\item{Robin, G. deQ.  1983.  In
\underline{Climate Record in Polar Ice Sheets} (Cambridge
University Press, New York), Chap. 4. }
\item{Royer, A., M. De Angelis, and J. R. Petit.  1983.
{\sl Climatic Change} {\bf 5}, 381-412. }
\item{Schott, D., E. D. Waddington, and C. F. Raymond.  1992.
{\sl Journal of Glaciology} {\bf 38}, 162-168. }
\item{Shoji, H., and Langway, C. C.  1982.  {\sl Nature} {\bf 298}
 (6874), 548-550. }
\item{Shoji, H., and Langway, C. C.  1987.
{\sl J. Phys. (Paris)}, {\bf 48}, Colloq. C1, 551-556
(Supplement au 3). }
\item{L. G. Thompson.  1977.  IAHS Publication {\bf 118}, 351-364. }
\item{Uchida, T.  1994.  Private communication. }
\item{Uchida, T., T. Hondoh, S. Mae, V. Ya. Lipenkov, and P.
Duval.  1994.  {\sl Journal of Glaciology} {\bf 40}, 79-86. }
\item{Uchida, T., T. Hondoh, S. Mae, P. Duval, and V. Ya.
Lipenkov.  In {\sl Fifth Inter. Symp. on Antarctic Glaciology},
to be published, 1994. }
\item{Weertman, J.  1968.  {\sl Journal of Geophysical Research}
 {\bf 73} (8),  2691-2700. }
\item{Young, N. W., M. Pourchet, V. M. Kotlyakov, P. A. Korolev,
and M. B. Dyugerov. 1982.  {\sl Ann. Glaciology} {\bf 3}, 333-338. }
\end{list}

\newpage

\section*{\normalsize \bf FIGURE CAPTIONS}
\bigskip

\newcounter{capnum}

\begin{list}{\arabic{capnum}.}{\usecounter{capnum}}
\item The AMANDA high-energy neutrino observatory.  The top four strings are
frozen into the ice at South Pole and are working.  The lower six strings will
be deployed at a greater depth to be discussed in this paper.
\item Depth as a function of age for various deep cores.  Curves are meant to
guide the eye.  Points are a sampling of more extensive data from the
references in Table 1.
\item Universal curve of dimensionless depth vs dimensionless age, with $a(t)$
= step function, reduced at 15-110 kyr BP (see text).
\item Models of depth as a function of age for South Pole ice.  Solid curve
(the AMANDA model) is for $a$ = 0.073 m/yr since 15 kyr BP and $a$ = 0.055 m/yr
before 15 kyr BP.  Dashed curve is the flowline model of McInnes and Radok
(1984).  Irregular curve is dust flux at Vostok after smoothing by a cubic
spline function (Petit et al., 1990).  The scale for dust flux is linear, with
largest peak corresponding to 7.3 $\times$ 10$^{-7}$ cm/yr.
\item Inverse scattering length as a function of depth for air bubbles.  The
data for Byrd and Vostok were taken from microscopic measurements of bubbles in
cores.  The AMANDA data for South Pole were based on light scattering of laser
pulses in situ.  The solid curve shows the effect of hydrostatic pressure on
bubble sizes, assuming all of the air is trapped in bubbles.  The dashed curve
shows the calculated dependence on depth due to conversion of bubbles into air
hydrate crystals (Price, 1994).
\item Temperature profiles for several sites in Greenland and Antarctica,
compared with pressure dissociation equilibria (converted to depths) for
nitrogen-clathrates and air (N$_2$ + O$_2$) hydrates.  In the hatched region
both bubbles and hydrate crystals should co-exist. The solid triangles indicate
depths at which air hydrate crystals are first observed to appear; the solid
squares indicate depths at which air bubbles have completely disappeared.
Arrows at solid squares for Dome C and Camp Century indicate lower limits on
depths for disappearance of bubbles.
\end{list}

\newpage

\centerline{Table 1. Data on Ice Cores and Drilling Sites}

\begin{center}
\begin{tiny}
\begin{tabular}{|l|c|c|c|c|c|c|c|c|c|c|}    \hline
  & Lateral & Height, $h$, & Depth, & Elevation & Surf. & Recent & Accum.
& Bubbles & Hydrates & Depth of \\
  & Flow & of flowrate & $H$, to & (m) & temp. & accum. rate & rate & obs. to
& observed & max. dust \\
Location & rate & transition & bedrock &  & ($^\circ$ C) & (m ice/yr) & 15-110
kyr
& disappear & to appear & conc. \\
  & (m/yr) & (m) & (m) & & & & BP & at: & at: & \\ \hline
Camp Century$^a$ & 3.3 & 430 & 1388 & 1890 & -24 & 0.38 & n.s. & $>$1388 m
& 1100 m & 1200 m \\
 & & & & & & & & & & \\
Milcent$^b$ & 48 & n.s. & 2340 & 2450 & -22 & 0.53 & n.s. & n.s. & n.s.
& n.s. \\
 & & & & & & & & & & \\
GISP (Dye 3)$^c$ & 12.3 & 300 & 2037 & 2479 & -19.6 & 0.49 & 0.25 & 1540
& 1092 & n.s. \\
 & & & & & & & & & & \\
GRIP (Summit)$^d$ & 0 & 1200 & 3029 & 3238 & -32 & 0.23 & 0.12 & 1350
& n.s. & n.s. \\
 & & & & & & & & & & \\
Byrd Station$^e$ & 12.8 & ? & 2164 & 1520 & -27.9 & 0.13 & 0.1 & 1100
& 727 & 1450 \\
 & & & & & & & & & & \\
Vostok$^f$ & 3 & 400 & 3700 & 3488 & -55.6 & 0.024 & 0.018 & 1250 & 500
& 420 \\
 & & & & & & & & & & \\
Dome C$^g$ & 0 & ? & 3700 & 3240 & -53.5 & 0.037 & 0.028 & $>$800 & n.s.
& 600 \\
 & & & & & & & & & & \\
South Pole & 8-10$^h$ & 400-800$^h$ & 2900 & 2835 & -51 & 0.073$^i$ & 0.055
& $>$1000; & predict & predict \\
 & & & & & & & & predict & $\sim$500 & $\sim$1000 \\
 & & & & & & & & $\sim$1450 & & \\
\hline
\end{tabular}
\end{tiny}
\end{center}

\bigskip

\begin{list}{}{}
\item a. Dansgaard and Johnsen (1969); Dansgaard and others (1982); Hammer and
others
(1978).
\item b. Hammer and others (1978).
\item c. Dahl-Jensen and Johnsen (1986); Dansgaard and others (1982).
\item d. Schott and others (1992); Dahl-Jensen (1989); Dansgaard and
others(1993).
\item e. Robin (1983); Lorius and others (1984).
\item f. Ritz (1989); Lorius and others (1985); Jouzel and others (1993); De
Angelis
and others (1984).
\item g. Lorius and others (1979); J.-R. Petit and others (1981).
\item h. Koci (1994).
\item i. Mosley-Thompson and Thompson (1982); Giovinetto (1964); Young and
others
(1982).
\end{list}

\newpage

\centerline{Table 2.  Dust Concentration in Cores at Depths for Holocene,
LGM, and 40 kyr BP}

\begin{center}
\begin{tabular}{|l|c|c|c|c|} \hline
Location & Holocene & LGM & 40 kyr BP & Ref. \\ \hline
Camp Cent. & 2400 & 2 $\times$ 10$^5$ & not studied & a \\
Summit & 1 $\times$ 10$^4$ & not studied & not studied & b \\
Byrd & 1000 & 1 $\times$ 10$^4$ & 1500/ml & a \\
Dome C & 4200 & 7 $\times$ 10$^4$ & 1.1 $\times$ 10$^4$ $^\star$ & c \\
Vostok & 5650 & 6.5 $\times$ 10$^4$ & 1.6 $\times$ 10$^4$ & d \\
South Pole & 1450 & predict 1.9 $\times$ 10$^4$ & predict 3350 & e \\
\hline
\end{tabular}
\end{center}

$^\star$Measurement made near bottom of core, at $\sim$ 30 kyr BP.

(a) L.G. Thompson (1977); (b) Palais {\sl et al.} (1992); (c) Petit
{\sl et al.} (1981); (d) De Angelis {\sl et al.} (1984); (e) Gayley and
Ram (1985); E. Mosley-Thompson (1994).

\bigskip

\bigskip

\centerline{Table 3.  Size Distribution of Dust in a South Pole Core
at Depth of 100 - 349 m}

\begin{center}
\begin{tabular}{|c|c|c|} \hline
Diameter ($\mu$m) & No./ml $^\star$ & Area/ml \\ \hline
0.1 - 0.4 & 6 $\times$ 10$^4$ & 1.7 $\times$ 10$^{-5}$ cm$^{-1}$ \\
0.4 - 0.8 & 8200 & 2 $\times$ 10$^{-5}$ \\
0.81 - 1.0 & 885 & 6 $\times$ 10$^{-6}$ \\
1.01 - 1.25 & 578 & 6 $\times$ 10$^{-6}$ \\
1.26 - 1.59 & 343 & 5.5 $\times$ 10$^{-6}$ \\
1.6 - 2.0 & 190 & 5 $\times$ 10$^{-6}$ \\
$\geq$ 2.0 & 235 & 1.2 $\times$ 10$^{-5}$ \\ \hline
$\geq$ 0.1 & 7 $\times$ 10$^4$ & 7 $\times$ 10$^{-5}$ \\
\hline
\end{tabular}
\end{center}

$^\star$Gayley and Ram (1985); Mosley-Thompson (1994).

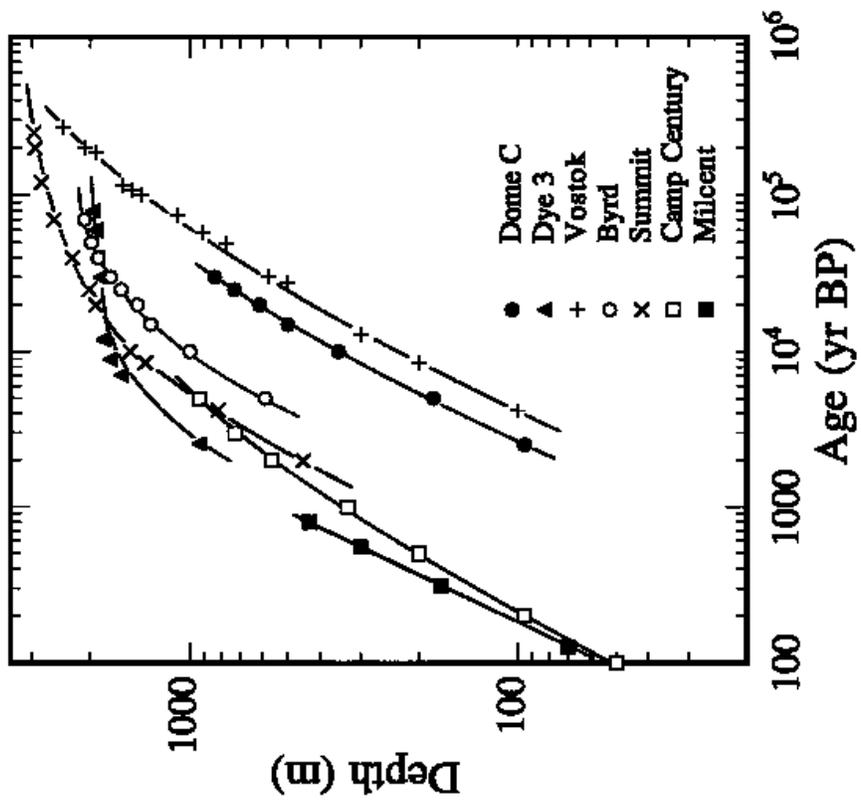
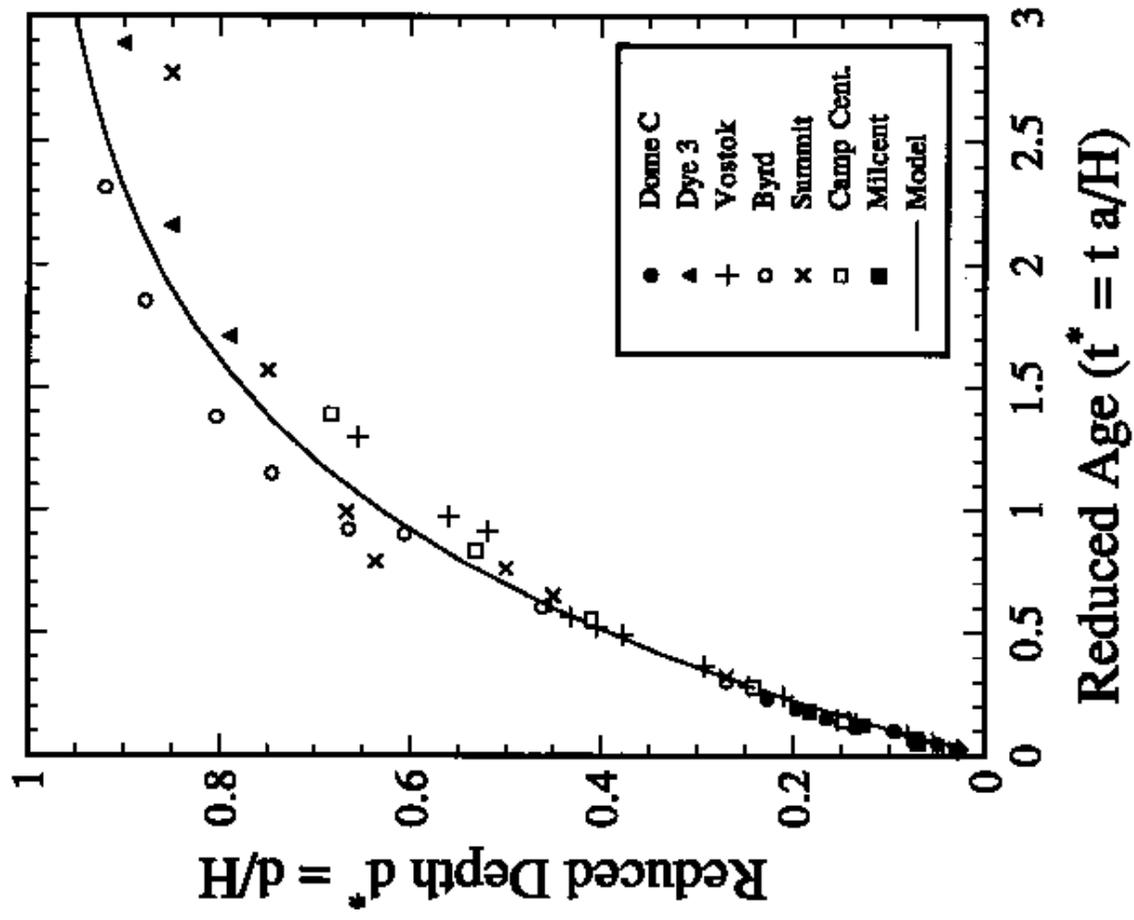
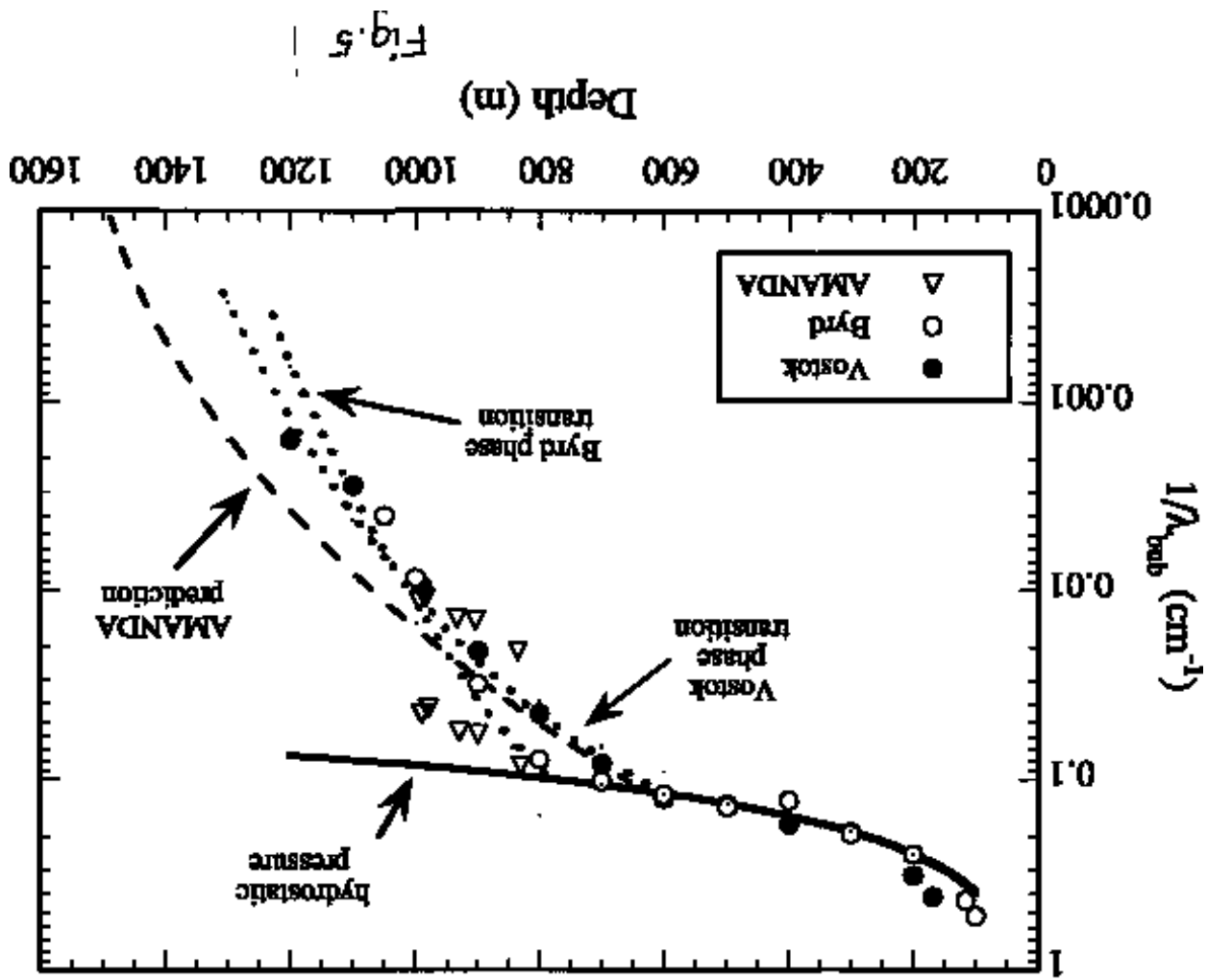
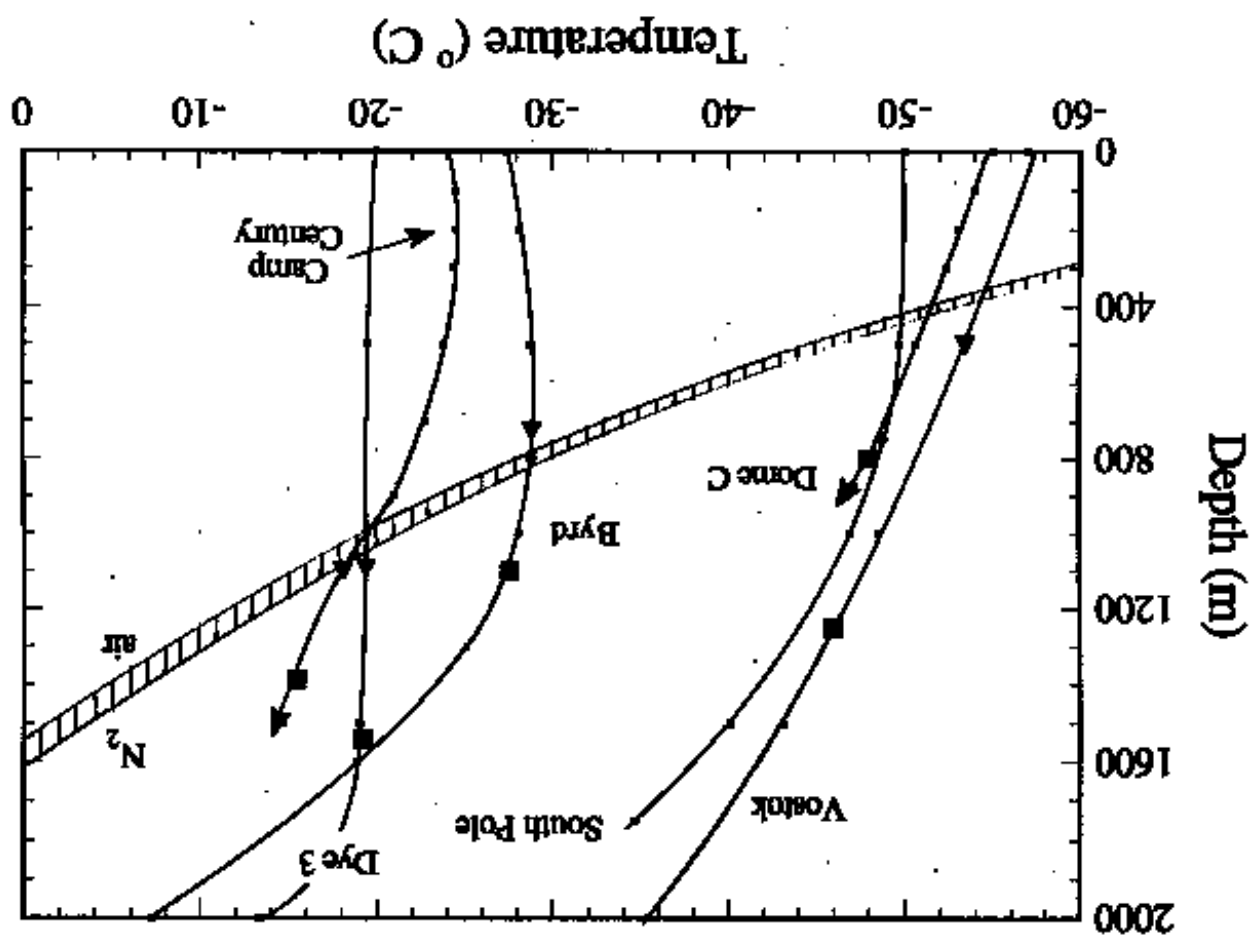
\end{document}